\documentclass[letterpaper,twocolumn,prl,
aps,showpacs,groupedaddress,floatfix]{revtex4-2}
\usepackage[latin1]{inputenc}
\usepackage{bm}
\usepackage[usenames]{color}
\usepackage{multirow}
\usepackage{amssymb}
\usepackage{amsbsy}
\usepackage{mathtools}
\usepackage{amsmath}
\usepackage{stmaryrd}
\usepackage{graphicx}
\usepackage{epsfig}
\usepackage{placeins}
\usepackage{ulem}
\usepackage{bbold}
\usepackage{braket}
\usepackage{blindtext}
\usepackage{nicefrac}
\usepackage[colorlinks,linkcolor=blue,citecolor=blue,urlcolor=blue]{hyperref}

\makeatletter

\begin{document}
\title{Anomalous hydrodynamics in a class of scarred frustration-free 
Hamiltonians}

\author{Jonas Richter}
\email{j.richter@ucl.ac.uk}
\affiliation{Department of Physics and Astronomy, University College London, 
Gower Street, London WC1E 6BT, UK}

\author{Arijeet Pal}
\affiliation{Department of Physics and Astronomy, University College London, 
Gower Street, London WC1E 6BT, UK}

\date{\today}

\begin{abstract}

Atypical eigenstates in the form of quantum scars and fragmentation of Hilbert 
space due to conservation laws provide obstructions to thermalization in the 
absence of disorder. In certain models with dipole and $U(1)$ conservation, the 
fragmentation results in subdiffusive transport. In this paper we study the 
interplay between scarring and weak fragmentation giving rise to anomalous 
hydrodynamics  in a class of one-dimensional spin-$1$ frustration-free 
projector 
Hamiltonians, known as deformed Motzkin chain. The ground states and low-lying 
excitations of these chains exhibit large entanglement and critical slowdown.   
We show that at high energies the 
particular form of the  
projectors causes the emergence of disjoint Krylov subspaces for 
open boundary 
conditions, with an exact quantum scar being embedded in each subspace, leading 
to slow growth of 
entanglement and localized dynamics for 
specific 
out-of-equilibrium initial states.
Furthermore, focusing on infinite temperature, we unveil
that spin transport is subdiffusive, which we 
corroborate by simulations of constrained stochastic 
cellular automaton circuits. Compared to dipole moment conserving systems, the  
deformed Motzkin chain appears to belong to a 
different universality class with distinct dynamical transport exponent and 
only polynomially many Krylov subspaces.

\end{abstract}

\maketitle

{\it Introduction.--}
Unraveling the intricate dynamics of isolated many-body quantum systems has 
attracted a vast amount of interest 
in recent years \cite{polkovnikov2011, gogolin2016, dalessio2016, 
borgonovi2016, Mori2018}. 
In this context, transport processes represent arguably one of 
the most generic nonequilibrium situations and the common 
expectation is that hydrodynamics emerges naturally from the underlying unitary 
time evolution \cite{Bertini2021, Khemani2018}.
The emergence of a variety of universal hydrodynamics and their relevance to  
transport coefficients are actively pursued theoretically with potential for 
utility in near-term quantum devices 
\cite{Bertini2021, Khemani2018, Ye2020, Richter2021}.
Enormous experimental efforts  
have been undertaken to study quantum transport in various platforms, including 
mesoscopic and solid-state settings as well as cold-atom quantum simulators 
(see e.g., \cite{DasSarma2011, Hess2019, 
Scheie2020, Hild2014, Jepsen2020}), remarkably allowing to observe even 
anomalous types of hydrodynamics  \cite{Wei2021, Joshi2021}.

While most quantum systems relax to thermal equilibrium, as explained by the 
eigenstate thermalization hypothesis (ETH) \cite{deutsch1991, 
srednicki1994, rigol2005} and numerically confirmed for a variety of models 
(e.g., 
\cite{dalessio2016, steinigeweg2013, beugeling2014, kim2014, Torres-Herrera2014,
Mondaini2016, jansen2019, LeBlond2019, Brenes2020, Richter2020}), several 
counterexamples to this paradigm have been identified, with integrable and 
many-body localized systems being prime examples \cite{essler2016, 
nandkishore2015, Abanin2019}. Moreover, studies of the so-called 
PXP model revealed that 
also weaker violations of the ETH are possible, where rare nonthermal states 
coexist with thermal eigenstates at the same energy density 
\cite{Bernien2017, Turner2018}, now usually referred to as quantum many-body 
scars 
\cite{Turner2018, Moudgalya2018, Moudgalya2018_2, Khemani2019, Choi2019, 
Lin2019, Serbyn2020}. By now, quantum scars 
have been found in various models \cite{Turner2018, 
Moudgalya2018, Moudgalya2018_2, Khemani2019, Choi2019, 
Lin2019, Serbyn2020, Schecter2019, 
Iadecola2019, Ok2019, James2019, Wildeboer2020, Lee2020, McClarty2020, 
Kuno2020, Zhao2020, Surace2020, vanVoorden2020, 
Pilatowsky-Cameo2021, Jeyaretnam2021, Banerjee2021}, and tailored 
embedding procedures further allow to place nonthermal eigenstates into the 
spectrum 
of chaotic many-body Hamiltonians \cite{Shiraishi2017, 
Shiraishi2019}.

Building on insights from fractonic systems 
\cite{Chamon2005, Pretko2018, Nandkishore2019}, the phenomenon of 
Hilbert-space fragmentation provides 
yet another mechanism to break ergodicity \cite{Pai2019, Sala2020, Khemani2020, 
Moudgalya2019}. Hilbert-space fragmentation occurs, for instance, in locally 
interacting models which in addition to a $U(1)$ charge also conserve the 
associated dipole moment, 
though other possibilities have been discussed as well \cite{DeTomasi2019, 
Yang2020, Li2021, Lee2020_2, Hahn2021}. In such  
cases, the 
Hilbert space splits into exponentially many disconnected blocks, 
often 
referred to as Krylov subspaces, despite states in different 
subspaces having the same symmetries. 
While some subspaces
might be integrable or localized, others can be chaotic 
\cite{Moudgalya2019, Yang2020, 
Herviou2020}.
Even within the thermalizing regimes of such models, the 
constraints on 
excitations, e.g., higher-order conservation laws, have 
implications on the dynamics and lead 
to subdiffusive transport \cite{Iaconis2019, 
Gromov2020, Feldmeier2020, Morningstar2020, Zhang2020, Iaconis2021, 
Moudgalya2021, Glorioso2021}, reminiscent 
of disordered models close to the many-body 
localization transition \cite{BarLev2015, Agarwal2015, Luitz2017}. The class 
of 
frustration-free Hamiltonians considered in this Letter similarly exhibits 
disjoint Krylov subspaces and subdiffusive hydrodynamics. The underlying 
mechanisms, however, will be distinct from those of the models 
mentioned above.

Another motivation for this paper is given by recent work on quantum 
many-body 
scars and Hilbert-space fragmentation in Fredkin chains 
\cite{Langlett2021}. 
The 
Fredkin model is a spin-$1/2$ chain, where the 
Hamiltonian is a sum over projectors and can be rewritten in the 
form of a dressed Heisenberg chain \cite{Salberger2017}, bearing resemblance to 
other kinetically constrained 
models \cite{Ritort2003, Lan2018, 
Pancotti2020}. 
While the model is nonintegrable in general, its degenerate ground-state 
manifold is 
known analytically \cite{Salberger2017}. 
In particular, as shown in \cite{Langlett2021}, the degenerate states can 
be moved to the center of the 
spectrum by generalizing the model \cite{Salberger2017_2, 
Adhikari2020},  
with each state belonging to a different Krylov subspace. 

Here, we consider a 
closely related class of models, known as deformed Motzkin chain 
\cite{Bravyi2012, Zhang2017, 
Movassagh2016, Levine2017, Movassagh2017, DellAnna2016, Babiero2017}.
While the ground-state 
properties of Motzkin chains have been explored in a series of works 
\cite{Bravyi2012, Zhang2017, 
Movassagh2016, Levine2017, Movassagh2017, DellAnna2016, Babiero2017, 
Chen2017, Chen2017_2, Sugino2018}, much less is known about the 
nature of thermalization and nonequilibrium dynamics. 
In this work, we show that the particular form of 
the Hamiltonian leads to an intriguing interplay of disconnected 
Krylov subspaces and exact quantum many-body scars, similar to 
\cite{Langlett2021}. As a main result, we unveil that the Motzkin chain 
exhibits subdiffusive hydrodynamics at infinite temperature, 
which 
we corroborate by simulations of 
suitable stochastic cellular automaton circuits \cite{Iaconis2019, 
Feldmeier2020, Morningstar2020, Medenjak2017, Gopalakrishnan2018}. Furthermore, 
we demonstrate that the scarred eigenstates 
lead to localized dynamics for specific out-of-equilibrium states and parameter 
regimes.
 
\begin{figure}[tb]
 \centering
 \includegraphics[width=0.9\columnwidth]{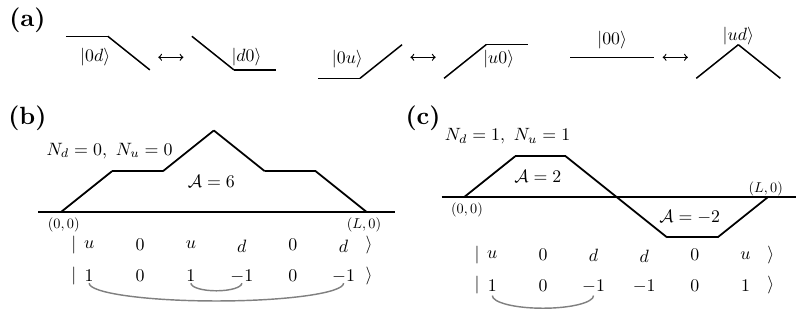}
 \caption{Identification of $\ket{1}$, $\ket{-1}$, 
$\ket{0}$ as $\ket{u}$, $\ket{d}$, $\ket{0}$, corresponding to 
\textit{up}, \textit{down}, and \textit{horizontal} moves on a plane. {\bf (a)} 
Local updates induced by the projectors of ${\cal H}_\nu$.  
{\bf [(b),(c)]} For OBC, the Hilbert space
splits into Krylov subspaces labeled by $N_d$ and 
$N_u$. Panel (b) shows 
example configuration with $N_d = N_u = 0$. Paired 
spins 
are indicated by arcs. Panel (c) shows a configuration with $N_d = N_u 
= 1$. The 
area ${\cal 
A}$ determines the weight of 
the basis state within $\ket{S_\nu}$, 
see Eq.~\eqref{Eq::AreaState}. }
 \label{Fig1}
\end{figure}

{\it The model.--}
We consider a class of spin-$1$ projector Hamiltonians, ${\cal H}_\nu = 
\sum_{\ell} \Pi_{\ell,\ell+1}(\nu)$,  known as deformed 
Motzkin chain \cite{Bravyi2012, Zhang2017, 
Movassagh2016, Levine2017, Movassagh2017, DellAnna2016, Babiero2017, 
Chen2017, Chen2017_2, Sugino2018},
 \begin{equation}\label{Eq::DeformedProjec}
 \Pi_{\ell,\ell+1}(\nu) = c_1\ket{D_\nu}\hspace{-0.08cm}\bra{D_\nu} +  
c_2\ket{U_\nu}\hspace{-0.08cm}\bra{U_\nu} + 
c_3\ket{V_\nu}\hspace{-0.08cm}\bra{V_\nu}\ , 
\end{equation}
where $c_1$, $c_2$ and $c_3$ are real-valued 
coefficients, $\nu\geq0$ is a deformation parameter,
and the terms $\ket{\cdot}\bra{\cdot}$ 
are given by $\ket{D} = (\ket{0d} - 
\nu\ket{d0})/\sqrt{1+\nu^2}$, $\ket{U} = (\ket{u0} 
- \nu\ket{0u})/\sqrt{1+\nu^2}$, $\ket{V} = (\ket{ud} - 
\nu\ket{00})/\sqrt{1+\nu^2}$ and should 
be understood as acting on two neighboring
sites $\ell$ and $\ell+1$. 
We adopt the convention to denote the three eigenstates of 
a local spin-$1$ operator $S_\ell^z$ as $\ket{u} 
\equiv \ket{+1}$, $\ket{d} \equiv \ket{-1}$ and $\ket{0}$, where $\ket{u}$ 
(``\textit{up}''), $\ket{d}$ 
(``\textit{down}''), and $\ket{0}$ are interpreted as the moves $(x,y) 
\to (x+1,y+1)$, $(x,y) \to 
(x+1,y-1)$, and $(x,y) \to 
(x+1,y)$ on a two-dimensional plane \cite{Movassagh2016}, see Fig.\ \ref{Fig1}.
The terms $\ket{\cdot}\bra{\cdot}$ in Eq.\ 
\eqref{Eq::DeformedProjec} have eigenvalues $0$ and $1$ such that ${\cal 
H}_\nu$ has a positive-semidefinite spectrum if all $c_i \geq 0$. ${\cal 
H}_\nu$ 
has a $U(1)$ symmetry, such that $S^z = \sum_\ell 
S_\ell^z$ is 
conserved.
Written in terms of usual spin-$1$ operators, ${\cal H}_\nu$ 
takes on a bilinear-biquadratic form \cite{Movassagh2017, Chen2017, 
Chen2017_2}. 

For a spin 
configuration on $L$ 
sites, the identification 
of spins as moves leads to a ``random walk''. 
In the $S^z = 0$ sector, these walks start at $(0,0)$ and end at 
$(L,0)$, see Figs.\ \ref{Fig1}~(b) and (c). For open 
boundary conditions (OBC), 
an important concept is then the 
distinction between {\it paired} and {\it unpaired} moves \cite{Bravyi2012}. An 
up move is called unpaired if there is no matching down move further to the 
right in the chain, and a down move is unpaired if there is no matching up move 
further to the left. Given a configuration with no unpaired moves, 
the height profile never crosses the horizon [Fig.\ 
\ref{Fig1}~(b)].  
Such walks in the upper half-plane are referred to as Motzkin paths, 
giving rise to 
the name of the model. 
\begin{figure}[tb]
 \centering
 \includegraphics[width=1\columnwidth]{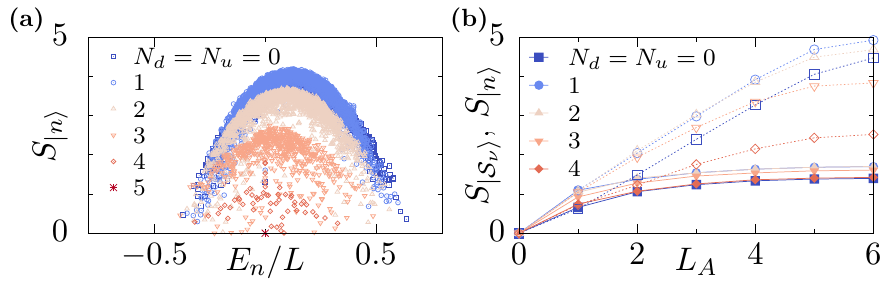}
 \caption{{\bf (a)} Eigenstate entanglement $S_{\ket{n}}$ for OBC, $L = 10$, 
and $\nu = 1$, labeled according to their Krylov subspace. 
\textbf{(b)} $S_{\ket{{\cal S}_\nu}}$ (filled, solid) at $\nu = 
1$ versus 
subsystem size $L_A$ for $L = 12$ 
and different ${\cal K}_{du}$ with $N_d = N_u$. As a 
comparison, the entanglement $S_{\ket{n}}$ (open, dashed) of an eigenstate 
directly adjacent to~$\ket{{\cal 
S}_\nu}$ is shown. We have $c_1 = c_3 = 1$, and $c_2 = -1$.
}
 \label{Fig2}
\end{figure}

{\it Disconnected Krylov subspaces.--} \label{Sec::Model_HSF}
In the case of OBC, the Hilbert space of ${\cal H}_\nu$ splits into Krylov 
subspaces due to the 
interplay of the boundary conditions and the action of the projectors on 
neighboring spins, cf.\ Fig.\ 
\ref{Fig1}~(a). The subspaces can be understood as 
equivalence classes, 
where each spin 
configuration is equivalent to a specific root state $\ket{\psi_{du}}$ 
\cite{Bravyi2012}. 
Given an arbitrary configuration, $\ket{\psi_{du}}$ can 
be defined as follows. First, identify {\it pairs} of up and down spins, where 
the spins  
forming a pair do not have to be 
nearest neighbors, cf. Figs.\ \ref{Fig1}~(b) and (c). Secondly, flip both 
spins to the $\ket{0}$ state and move the zeros to the center, which 
eventually yields \cite{Bravyi2012},   
\begin{equation}\label{Eq::Root}
\ket{\psi_{du}} = \ket{\underbrace{dd\cdots 
dd}_{N_d}\underbrace{00\cdots00}_{L-N_d-N_u}\underbrace{uu\cdots uu}_{N_u}}\ , 
\end{equation}
where $N_d$ and $N_u$ denote the numbers of unpaired down or up 
moves. Given $\ket{\psi_{du}}$, its corresponding 
Krylov 
subspace ${\cal 
K}_{du} = {\cal K}({\cal H}_\nu,\ket{\psi_{du}})$ 
follows as
 ${\cal 
K}_{du} = \text{span}\lbrace \ket{\psi_{du}}, 
{\cal 
H}_\nu\ket{\psi_{du}}, {\cal H}^2_\nu\ket{\psi_{du}}, \dots\rbrace$.
In particular, two spin configurations which correspond to different 
$\ket{\psi_{du}}$ cannot 
be transformed into each other by the action of ${\cal H}_\nu$.
As an example, consider $\ket{\psi_1} = \ket{u 
\cdots ud\cdots d}$ and $\ket{\psi_2} = \ket{d \cdots du\cdots u}$, 
which both have $S^z = 0$. However, while 
$\ket{\psi_1}$ belongs to ${\cal K}_{00}$ (i.e., 
it 
is equivalent to $\ket{0\cdots 0}$), $\ket{\psi_2}$ belongs to 
${\cal K}_{\nicefrac{L}{2} \nicefrac{L}{2}}$. In fact, $\ket{\psi_2}$ 
is an exact eigenstate of ${\cal H}_\nu$, i.e., it spans a subspace of 
dimension one.
Apparently the degree of ``Hilbert-space 
fragmentation'' in the Motzkin chain
is weaker compared to, e.g., models with charge and dipole conservation, 
which exhibit exponentially many subspaces \cite{Pai2019, Sala2020, 
Khemani2020}. 
For instance, in the $S^z = 0$ sector, there are only 
$L/2+1$ separate ${\cal K}_{du}$ labeled by $0 \leq N_d = N_u \leq L/2$, i.e., 
the total number of subspaces grows only polynomially with $L$. 
An expression for the dimension ${\cal D}_{du}$ of each ${\cal 
K}_{du}$ can 
be derived combinatorially 
\cite{NoteDimension}.
In particular, for ${\cal K}_{du}$ with small $N_d + N_u$, ${\cal D}_{du}$ 
is expected to grow exponentially with $L$. At the 
same time, for any finite $L$, there always exist ${\cal K}_{du}$ with ${\cal 
D}_{du} = 1$ (namely when $N_d + N_u = L$), as well as small 
subspaces with ${\cal D}_{du} \propto L$.  

For subspaces 
with large ${\cal D}_{du}$, thermalization is expected to occur.
This is visualized in Fig.\ \ref{Fig2}~(a) in terms of the eigenstate 
entanglement 
entropy $S_{\ket{n}} = -\text{Tr}[\rho_A \ln \rho_A]$, where $\rho_A = 
\text{Tr}_B 
\lbrace \ket{n}\bra{n}\rbrace$ is the reduced density matrix for a half-chain 
bipartition. 
While the overall distribution of $S_{\ket{n}}$ is rather broad, it 
looks thermal when focusing on individual ${\cal K}_{du}$ with small 
$N_d,N_u$. 
At the same time, the low values of $S_{\ket{n}}$ in the center of the 
spectrum mostly belong to ${\cal K}_{du}$ with large $N_d,N_u$, 
where the maximally achievable entanglement is limited due to small 
${\cal 
D}_{du}$. 
Moreover, as shown in \cite{SuppMat}, individual ${\cal K}_{du}$ indeed 
exhibit 
chaotic energy-level statistics and most eigenstates follow the 
ETH.

{\it Exact quantum many-body 
scars.--} Despite ${\cal H}_\nu$ being nonintegrable and chaotic, a 
number 
of eigenstates $\ket{{\cal S}_\nu}$ can be constructed 
combinatorially \cite{Bravyi2012, Zhang2017, Movassagh2016}.
In this context, the key quantity is 
the 
area ${\cal A}_k$ enclosed by the height profile of a given spin configuration 
$\ket{k}$, where areas below the horizon 
contribute negatively, cf.\ Figs.~\ref{Fig1}~(b) and (c). Within each ${\cal 
K}_{du}$, $\ket{{\cal S}_\nu}$ is 
then given by the area-weighted superposition 
\cite{Bravyi2012, Zhang2017, 
Movassagh2016} (see also \cite{SuppMat}), 
\begin{equation}\label{Eq::AreaState}
 \ket{{\cal S}_\nu} = \frac{1}{\sqrt{M_\nu^\prime}} \sum_{k=1}^{{\cal D}_{du}} 
\nu^{{\cal A}_k} \ket{k}= \frac{1}{\sqrt{M_\nu}}\sum_{k=1}^{{\cal D}_{du}} 
\nu^{-{\cal P}_k} 
\ket{k}\ , 
\end{equation}
where the sum runs over all ${\cal D}_{du}$ basis states $\ket{k}$, ${\cal P} = 
\sum_{\ell = 1}^L \ell S_\ell^z$ is the dipole 
operator with
${\cal P}_k = \bra{k}{\cal P}\ket{k}$, and $M'_\nu$ and $M_\nu$ 
ensure normalization.
The states $\ket{{\cal S}_\nu}$ have exactly zero energy as they are 
annihilated by all projectors in Eq.\ \eqref{Eq::DeformedProjec} 
\cite{Bravyi2012, 
Zhang2017, 
Movassagh2016}.
According to Eq.\ \eqref{Eq::AreaState}, $\ket{{\cal S}_\nu}$ is dominated by 
$\ket{k}$ with large positive ${\cal P}_k$ if $\nu < 1$. In 
contrast, for $\nu > 1$, $\ket{k}$ with large 
negative ${\cal P}_k$ dominate. At $\nu = 
1$, $\ket{{\cal S}_\nu}$ is an
equal-weight superposition of all states in ${\cal K}_{du}$, reminiscent 
of the Rokhsar-Kivelson ground state in quantum dimer models 
\cite{Rokhsar1988}.  

By choosing suitable $c_i$ in Eq.\ 
\eqref{Eq::DeformedProjec}, 
the $\ket{{\cal S}_\nu}$ can be shifted close to the center of the 
spectrum \cite{NoteCi}, where they act as quantum many-body scars 
due to their subvolume-law entanglement \cite{Bravyi2012, Zhang2017, 
Movassagh2016}, similar to other examples of 
frustration-free ground states being embedded by deforming the underlying 
model \cite{Ok2019, Wildeboer2020, Lee2020}.
The nonthermal nature of the $\ket{{\cal S}_\nu}$ is emphasized in 
Fig.~\ref{Fig2}~(b), where $S_{\ket{{\cal S}_\nu}}$ is shown versus subsystem 
size 
$L_A$ 
for different ${\cal K}_{du}$. 
In particular, $S_{\ket{{\cal S}_\nu}}$ is compared to the 
entanglement of an eigenstate directly adjacent 
to $\ket{{\cal S}_\nu}$, demonstrating that typical eigenstates are extensively 
entangled whereas
$\ket{{\cal S}_\nu}$ is not. As shown in \cite{SuppMat}, $\ket{{\cal S}_\nu}$ 
also violates the ETH by yielding atypical expectation values for local 
operators. 

While the construction of ${\cal K}_{du}$ as in Eq.\ \eqref{Eq::Root} does not 
apply to periodic boundary conditions (PBC), we note that quantum scars appear 
to exist also for PBC \cite{SuppMat}.
\begin{figure}[tb]
 \centering
 \includegraphics[width=1\columnwidth]{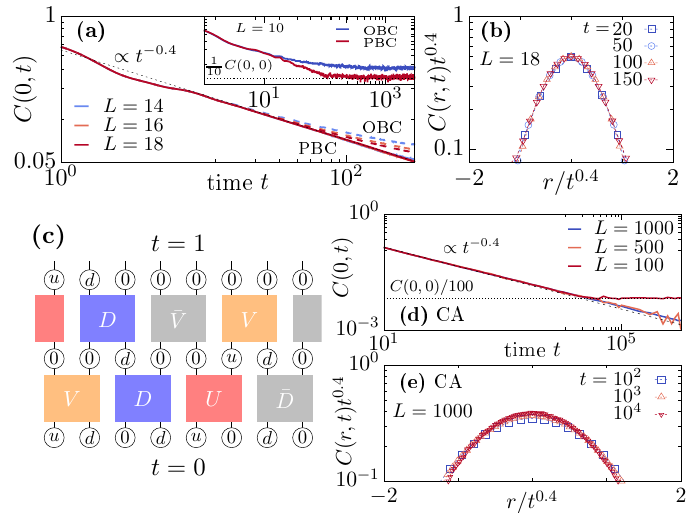}
 \caption{\textbf{(a)} $C(0,t)$ for PBC (solid) and 
OBC (dashed) at $\nu = 
1$ and $L = 14,16,18$. A 
power law $\propto t^{-1/z}$ with $z = 5/2$ is shown for comparison (dotted). 
The 
inset shows data for $L = 10$ up to longer times. 
\textbf{(b)} $C(r,t)t^{1/z}$ versus $r/t^{1/z}$ at fixed 
times. We have 
$c_1 = c_3 = 1$ and 
$c_2 = -1$ in all cases. 
\textbf{(c)} 
Exemplary time step in the cellular automaton (CA) 
circuit, consisting of two layers of two-site updates. Given 
a particular configuration of two sites, one of the 
updates $D$, $U$, or $V$ is chosen, while with probability $1/2$, we 
instead apply $\bar{D}$, $\bar{U}$, or $\bar{V}$, leaving the spin 
configuration unchanged (see \cite{SuppMat} for more details).
[\textbf{(d),(e)}] Analogous data as in panels (a) and (b), 
but now obtained by CA circuits for larger $L$.
}
 \label{Fig3}
\end{figure}

{\it Anomalous hydrodynamics.--}
We probe the transport properties of ${\cal H}_\nu$ in terms of the 
infinite-temperature correlation function $C(r,t)$, 
\begin{equation}\label{Eq::SpinCorr}
 C(r,t) = \text{Tr}[S_{\ell+r}^z(t)S_{\ell}^z]/3^L\ , 
\end{equation}
where $S_{\ell+r}^z(t) = e^{i{\cal H}t}S_{\ell+r}^ze^{-i{\cal H}t}$, and $r$ 
is the distance between the two sites \cite{NotePos}. 
In case of diffusion, $C(r,t)$ takes on a Gaussian shape with a standard 
deviation $\sigma(t) \propto 
t^{1/z}$ with $z = 2$ \cite{Bertini2021, Richter2019}. Correspondingly, the 
autocorrelation function $C(r = 0,t)$ acquires a hydrodynamic tail, 
$C(0,t) \propto t^{-1/z}$. For a thermalizing system, one expects 
a uniform distribution at long times, $C(r,t\to \infty) \to C_\text{eq}$, where 
 $C_\text{eq} = 
C(0,0)/L$~\cite{Bertini2021}.
We exploit quantum typicality 
\cite{SuppMat, Jin2021, Heitmann2020} to simulate $C(r,t)$ for spin-$1$ systems 
up to $L 
= 18$, beyond the range of full exact diagonalization. Focusing on 
$\nu = 1$, we find that $C(0,t) \propto t^{-1/z}$ 
with $z \approx 5/2$ (similar to \cite{Glorioso2021}),
suggesting that spin transport in the Motzkin 
chain is not diffusive but subdiffusive instead, both for PBC and OBC [Fig.\ 
\ref{Fig3}~(a)]. In the 
latter case, the power law persists on a shorter time scale 
as $C(0,t)$ saturates to a higher long-time value 
$C(0,t\to\infty) > 
C_\text{eq}$ [inset of Fig.\ \ref{Fig3}~(a)] due to the disjoint ${\cal 
K}_{du}$. We expect this difference between PBC and OBC to disappear in the 
thermodynamic limit $L \to \infty$, where 
the exponentially large ${\cal 
K}_{du}$ dominate.
Subdiffusive spin transport is further substantiated
in Fig.~\ref{Fig3}~(b), where  
the correlations $C(r,t)$ for 
different $t$ nicely collapse onto each other if the data and $r$ 
are rescaled with $t^{1/z}$.
We note that the observed value of $z$ is distinct from that found in 
dipole-conserving systems, where $z = 4$~\cite{Sala2020,Feldmeier2020}. 

Intuitively, the occurrence of subdiffusion can be understood by considering 
the updates of local spin configurations induced by 
${\cal H}_\nu$, cf.\ Fig.\ \ref{Fig1}~(a). As there are no matrix elements 
connecting $\ket{du} 
\leftrightarrow \ket{00}$ or $\ket{du} \leftrightarrow \ket{ud}$, 
configurations $\ket{du}$ act as bottlenecks. Particularly,
extended regions of the form $\ket{\cdots ddduuu \cdots}$ will 
slow down the dynamics. This argument can also be stated more formally by 
inspecting 
the spin-current 
operator of ${\cal H}_\nu$, see \cite{SuppMat}. 
While we cannot provide a full hydrodynamic theory, we here proceed by 
constructing a stochastic cellular 
automaton (CA) circuit, see 
Fig.\ \ref{Fig3}~(c) 
and \cite{SuppMat}, which mimics the terms appearing in ${\cal H}_\nu$ and 
allows to access large 
systems and 
long times \cite{Iaconis2019, 
Feldmeier2020, Morningstar2020, Medenjak2017, Gopalakrishnan2018}. The 
so-obtained data for 
$L \leq 10^3$ and $t \leq 10^6$ in Figs.~\ref{Fig3}~(d) and (e) corroborate 
our findings 
of anomalous hydrodynamics with $z \approx 5/2$ at infinite temperature. 
(Our CA data for large $L$ 
and long $t$ is also consistent with $z \approx 8/3$ \cite{Singh2021}.)
Putting 
these results into perspective, we 
note that subdiffusive dynamics in  
Motzkin chains \cite{Chen2017,Chen2017_2,DellAnna2016} (and related Fredkin 
models 
\cite{Adhikari2020_2}) has been observed before at low temperatures 
by analyzing the  
scaling of 
low-lying energy gaps, where a 
 slightly larger $z$ was found.  
In this context, we note that the dynamical exponent $z$ in 
certain constrained chaotic models consisting of Floquet random unitary 
circuits 
can be related to the scaling of the low energy gap of Rokhsar-Kivelson type 
Hamiltonians using classical Markov circuits \cite{Henley2004, Castelnovo2005,  
Singh2021, Moudgalya2021}, which has partially motivated our usage of CA 
circuits.

\begin{figure}[tb]
 \centering
 \includegraphics[width=1\columnwidth]{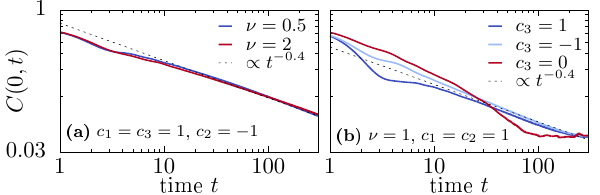}
 \caption{\textbf{(a)} $C(0,t)$ at $\nu = 0.5,2$ for $c_1 = c_3 = 1$, $c_2 = 
-1$. A 
power law $\propto t^{-0.4}$ is shown for 
comparison. \textbf{(b)} $C(0,t)$ at $\nu = 1$ for $c_3 = -1,0,1$ and $c_1 = 
c_2 = 1$. 
We 
have $L = 16$ and PBC in all cases.} 
 \label{Fig4}
\end{figure}

While we have focused on $\nu = 1$ in Fig.\ \ref{Fig3}, we 
stress that the 
occurrence of high-temperature subdiffusion seems
robust 
for a wider range of parameters. This is demonstrated in Fig.\ \ref{Fig4}, 
where 
$C(0,t)\propto t^{-1/z}$ both for $\nu = 0.5,2$, as well 
as for $\nu = 1$ but different choices of $c_i$.
Only for $c_3 = 0$, the decay of $C(0,t)$ appears to be different, which 
can be 
explained by the fact that ${\cal H}_\nu$ becomes integrable in this 
limit \cite{Tong2020}.

As an aside, we note that the 
anomalous transport properties of   
${\cal H}_\nu$ also reflect themselves in an unusual growth of 
R\'enyi entropies $S_\alpha(t) = \ln 
\text{Tr}[\rho_A^\alpha]/(1-\alpha)$, $\rho_A = 
\text{Tr}_B \ket{\psi(t)}\bra{\psi(t)}$, which were argued to grow 
subballistically for $\alpha > 1$ \cite{Rakovszky2019, Huang2020}, see 
\cite{SuppMat} for details.

\begin{figure}[tb]
 \centering
 \includegraphics[width=0.85\columnwidth]{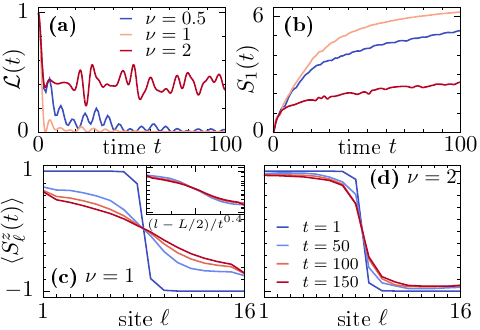}
 \caption{Dynamics of domain-wall state for $L = 16$ and OBC. 
[\textbf{(a),(b)}] ${\cal L}(t) = 
|\langle \psi(t)|\psi\rangle|^2$ and 
$S_1(t)$ for $\nu = 0.5,1,2$. [\textbf{(c),(d)}] 
$\langle 
S_\ell^z(t)\rangle$ at fixed 
$t$ for $\nu = 1,2$. Inset in (c) shows data at $t = 
50,100,150$ versus $(\ell-L/2)/t^{0.4}$.}
 \label{Fig5}
\end{figure}

{\it Initial-state dependence.--} While 
$C(r,t)$ represents a high-temperature average, studying quantum quenches 
with individual out-of-equilibrium states reveals the impact of the quantum 
scar $\ket{{\cal S}_\nu}$ on the dynamics. In particular, given its 
construction in Eq.\ \eqref{Eq::AreaState}, the dynamics 
can be tuned between different regimes depending on the 
deformation 
parameter $\nu$. We here exemplify this fact 
by considering a domain wall $\ket{\psi} = 
|u\dots ud \dots d \rangle$,
which is a
natural 
initial condition for quench dynamics \cite{Gobert2005, Ljubotina2017, 
Hauschild2016,
Medenjak2020}.
While $\ket{\psi}$ has zero energy density, $\bra{\psi}{\cal 
H}_\nu \ket{\psi}/L \to 0$, such that thermalization is expected, we note that 
in the picture of random-walks on a plane 
(Fig.\ \ref{Fig1}), $\ket{\psi}$ maximizes the area ${\cal 
A}$. According to the construction of $\ket{{\cal S}_\nu}$ in Eq.\ 
\eqref{Eq::AreaState}, $\ket{\psi}$ 
therefore contributes dominantly to $\ket{{\cal S}_\nu}$ if $\nu > 1$ 
(here $|\langle 
\psi|{\cal S}_\nu\rangle|^2 \approx 0.64$ for $\nu = 2$ and $L = 16$ 
\cite{NoteReimann}, in contrast to $|\langle 
\psi|{\cal S}_\nu\rangle|^2 = 1/{\cal D}_{du}^2$ for $\nu = 1$).
As a consequence, we find that
${\cal L}(t) = |\braket{\psi(t)|\psi}|^2$ decays  
quickly for $\nu = 0.5, 1$, while ${\cal L}(t)$  
oscillates around a finite value for $\nu = 2$ [Fig.\ \ref{Fig5}~(a)].
Likewise, the growth of the von Neumann entropy $S_1(t)$ \cite{NoteEntang} is 
significantly slower for $\nu = 2$ [Fig.\ \ref{Fig5}~(b)]. As shown in  
\cite{SuppMat}, there also exist initial states where dynamics is instead 
slower for $\nu < 1$ and faster for $\nu > 1$.

By tuning $\nu$ and thereby controlling its overlap with 
$\ket{{\cal S}_\nu}$, it is thus possible to obstruct thermalization of 
$\ket{\psi}$. This is emphasized even more in Figs.\ \ref{Fig5}~(c) and (d), 
where the spin profiles $\langle S_\ell(t) \rangle = 
\bra{\psi(t)}S_\ell^z\ket{\psi(t)}$ are shown at fixed times for $\nu = 1$ and 
$\nu = 2$. In particular, for $\nu = 2$, $\langle S_\ell(t)\rangle$ is 
found to remain localized even at long times. In contrast, for $\nu = 1$, 
$\ket{\psi}$ is not dominated by $\ket{{\cal S}_\nu}$ such that the domain wall 
melts away, albeit $\langle S_\ell(t) \rangle$ is still rather inhomogeneous 
even at $t = 150$. In fact, the profiles for different $t$ 
approximately collapse onto a 
single curve when plotted against 
$(l-L/2)/t^{1/z}$ [inset of Fig.\ \ref{Fig5}~(c)], i.e., 
consistent with the anomalous transport discussed 
above. We note that similar parameter-dependent melting of 
domain-wall  
states is known for other classes of models as well~\cite{Gobert2005, 
Medenjak2020}.

{\it Conclusion \& Outlook.--}
To summarize, we have studied a class of 
frustration-free Hamiltonians, 
where disjoint Krylov subspaces, anomalous hydrodynamics, and exact quantum 
many-body scars occur simultaneously. Compared to dipole-conserving or other 
fractonic models, the Motzkin chain appears to lie in a different 
``universality class'' featuring a distinct dynamical transport exponent $z 
\approx 5/2$ at infinite temperature and Hilbert-space fragmentation with 
only polynomially many subspaces. The quantum scars $\ket{{\cal 
S}_\nu}$ are similar to other embeddings of frustration-free 
ground states by deforming the 
underlying model \cite{Wildeboer2020, Lee2020, Ok2019}. Moreover, a similar 
construction of exact scars in individual Krylov 
subspaces has been recently presented for related Fredkin chains 
\cite{Langlett2021}.     

Regarding prospective directions of research, we note that while at present an 
analytical expression is known only for the states $\ket{{\cal 
S}_\nu}$, the data in Fig.\ \ref{Fig2} suggest 
that ${\cal H}_\nu$ hosts other low-entangled 
eigenstates beyond $\ket{{\cal 
S}_\nu}$. Approximating 
further nonthermal eigenstates, e.g., by devising a
spectrum generating algebra \cite{Mark2020, Moudgalya2020, 
Pakrouski2020, ODea2020} acting on 
$\ket{{\cal 
S}_\nu}$, might thus be an interesting attempt.  
Another extension is to study hydrodynamics at finite temperatures to connect 
our high-temperature results to the subdiffusive scaling of 
low-energy 
excitations \cite{Chen2017, Chen2017_2}, as well as to consider 
transport beyond half-filling, where CA circuits have already proven  
helpful \cite{Morningstar2020}.
Finally, the stability of the $\ket{{\cal 
S}_\nu}$ and, particularly, the persistence of anomalous hydrodynamics upon  
adding different perturbations to ${\cal H}_\nu$ is an open question.   

{\it Acknowledgements.--} 
We thank S.\ Moudgalya, B.\ Ware, and R.\ Vasseur for helpful 
comments.
This work was funded by the European Research 
Council 
(ERC) under the European Union's Horizon 2020 research and innovation programme
(Grant agreement No.\ 853368).

\appendix

\clearpage
\newpage

 
\setcounter{figure}{0}
\setcounter{equation}{0}
\renewcommand*{\citenumfont}[1]{S#1}
\renewcommand*{\bibnumfmt}[1]{[S#1]}
\renewcommand{\thefigure}{S\arabic{figure}}
\renewcommand{\theequation}{S\arabic{equation}}

\section*{Supplemental material}

\subsection{Krylov-space restricted thermalization}\label{Sec::Results_Therm}

\subsubsection{Periodic versus open boundary conditions} 

\begin{figure}[b]
 \centering
 \includegraphics[width = 0.9\columnwidth]{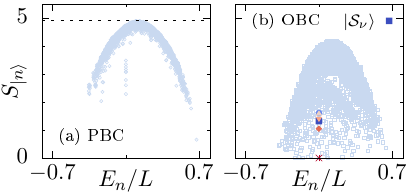}
 \caption{Eigenstate entanglement for (a) PBC and (b) OBC in the $S^z = 0$ 
sector. The horizontal dashed line in (a) indicates the entanglement of a 
random state. In (b), the exact quantum scars $\ket{{\cal S}_\nu}$ belonging 
to different Krylov subspaces are indicated by symbols. We have $L = 10$, $c_1 
= c_3 = 1$, and $c_2 = -1$ in all cases.}
 \label{FigS1}
\end{figure}

As discussed in the main text, the choice of OBC causes the emergence 
of disjoint Krylov subspaces ${\cal K}_{du}$. This fact can be exemplified by 
contrasting 
the entanglement entropies $S_{\ket{n}}$ of eigenstates of 
${\cal H}_\nu$ for PBC and OBC.
In the case of PBC [Fig.\ \ref{FigS1}~(a)], we find that 
$S_{\ket{n}}$ 
behaves as one would expect for a nonintegrable and thermalizing system, 
i.e., $S_{\ket{n}}$ takes on extensive values in the center of the spectrum 
which are similar to those of random states, while the 
entanglement towards the edges of the spectrum is lower. Interestingly,  
a degenerate set of 
zero-energy eigenstates, well separated from the band of thermal states, can be 
identified as well. Thus, similar to the 
states $\ket{{\cal S}_\nu}$ discussed in the main text, scarred eigenstates 
also exist for PBC. Specifically, since lattice momentum 
$\kappa$ is 
a good quantum number for PBC, these quantum scars then belong to sectors with 
different 
$\kappa$~\cite{Salberger2017S}. In contrast, in the case of OBC [Fig.\ 
\ref{FigS1}~(b)], the distribution of  $S_{\ket{n}}$ is rather broad due to 
the disjoint ${\cal K}_{du}$. Note that the data in Fig.\ \ref{FigS1}~(b) was 
already shown in Fig.\ 2~(a) of the main text. In Fig.\ 
\ref{FigS1}~(b) we now additionally highlight the comparatively low 
entanglement of the quantum many-body scars $\ket{{\cal S}_\nu}$ belonging to 
different ${\cal K}_{du}$.

\subsubsection{Level-spacing distribution}

\begin{figure}[b]
 \centering
 \includegraphics[width=0.9\columnwidth]{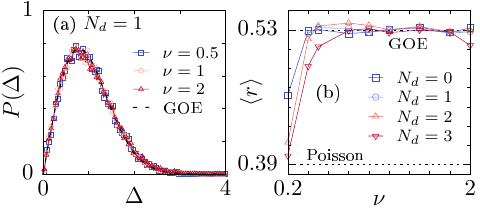}
 \caption{(a) Distribution $P(\Delta)$ of energy gaps in 
the Krylov space with 
$N_d = N_u = 1$ for $\nu = 0.5,1,2$. The dashed curve indicates 
the quantum-chaotic Wigner-Dyson distribution. Note that the correct 
extraction of 
$P(\Delta)$ requires an unfolding of the spectrum. (b)  $\langle 
r\rangle$ versus $\nu$ for Krylov spaces with different $N_d = N_u$. The 
dashed horizontal lines indicates the GOE and the Poissonian value 
respectively. 
We have $L = 12$, $c_1 = c_3 = 1$, and $c_2 = -1$ in all cases.}
 \label{FigS2}
\end{figure}

As the Hilbert space splits into Krylov spaces with a fixed number of 
unpaired spins $N_d, N_u$, chaos and thermalization has to be 
studied within each such subspace. A common diagnostic is the 
distribution $P(\Delta)$ of adjacent level spacings $\Delta_n = E_{n+1}-E_n$. 
In Fig.~\ref{FigS2}~(a), $P(\Delta)$ is shown for a fixed system size $L = 12$ 
in the largest Krylov subspace with $N_d = N_u = 1$. For all values of $\nu$ 
considered here, we find that $P(\Delta)$ accurately 
follows a Wigner-Dyson distribution, 
indicating the 
onset of quantum chaos \cite{dalessio2016S}. 

To study the statistics of energy levels for a wider range of 
$\nu$ and $N_d$, Fig.\ \ref{FigS2}~(b) shows the mean ratio $\langle r \rangle$ 
of adjacent level spacings \cite{Oganesyan2007S}, 
\begin{equation}
 \langle r \rangle = \frac{1}{N} \sum_n \frac{\text{min}\lbrace 
\Delta_n,\Delta_{n+1} \rbrace}{\text{max}\lbrace 
\Delta_n,\Delta_{n+1}\rbrace}\ , 
 \end{equation}
where the averaging is here performed over roughly $N=2{\cal D}_{du}/3$ of the 
eigenstates 
around the center of each Krylov space. For chaotic models, one expects 
$\langle 
r 
\rangle$ to be similar to the ratio of a random matrix, e.g., drawn from the 
Gaussian 
orthogonal ensemble (GOE), $\langle r\rangle_\text{GOE} \approx 0.53$. In 
contrast, for integrable or many-body localized models, the level spacing is 
Poissonian with $\langle r \rangle \approx 0.39$ \cite{dalessio2016S}. 

As shown in Fig.\ \ref{FigS2}~(b), we find $\langle r \rangle \approx 0.53$ 
for almost the whole range of $\nu \leq 2$ 
considered here (the agreement is slightly better for Krylov spaces with a 
larger dimension ${\cal D}_{du}$). Only if $\nu$ becomes too small, 
deviations from the random-matrix value appear, which can be explained by 
${\cal H}_\nu$ becoming entirely diagonal for $\nu \to 0$ with a  
highly degenerate spectrum.
We have spot-checked that $\langle r\rangle \approx 0.53$ also holds for the 
largest Krylov spaces in sectors with $S^z \neq 0$ (not 
shown here).
Since the dimensions ${\cal D}_{du}$ of Krylov spaces with large $N_d, 
N_u$ become too small to obtain good statistics, we 
refrain from searching in more detail for ${\cal K}_{du}$ that might exhibit 
Poissonian statistics. We note, however, that subspaces with large $N_d,N_u$
can indeed 
exhibit peculiar behavior. For instance, the subspace with $N_d = N_u = 
4$, depicted in Fig.\ 2~(a) of the main text, features a degenerate 
set of 
zero-energy eigenstates in addition to the exact state $\ket{{\cal S}_\nu}$.

\subsubsection{Validity of eigenstate thermalization hypothesis (ETH)}

\begin{figure}[tb]
 \centering
 \includegraphics[width=1\columnwidth]{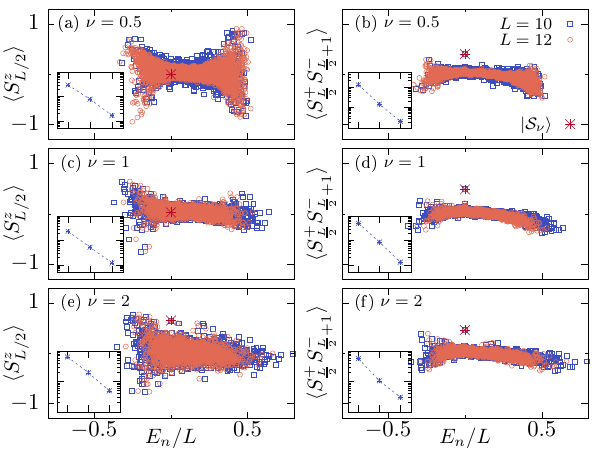}
 \caption{Eigenstate expectation values $\langle \cdot \rangle = 
\bra{n}\cdot \ket{n}$ versus energy density for the two operators 
 $S_{L/2}^z$ (left column) 
and $S_{L/2}^+ S_{L/2+1}^-$ (right column). Data is shown for $\nu = 0.5,1,2$ 
(top to bottom) and two 
different system sizes $L = 10,12$ in the Krylov space with $N_d = N_u = 0$. 
The 
expectation values with respect to the exact scar eigenstate $\ket{{\cal 
S}_\nu}$ are highlighted by an asterisk. The insets show the variances of 
the $\bra{n}\cdot \ket{n}$, evaluated within a narrow 
energy window in the center of the spectrum (excluding the scar $\ket{{\cal 
S}_\nu}$), versus system sizes $L = 8,10,12$. For all cases considered, the 
variances decrease approximately exponentially with $L$.}
 \label{FigS3}
\end{figure}

According to the ETH, the diagonal matrix 
elements of physical operators 
written in the eigenbasis of chaotic Hamiltonians should be a smooth function 
of energy \cite{dalessio2016S}. In Fig.\ \ref{FigS3}, we test 
the 
ETH for two local operators defined 
in the center of the chain, 
\begin{equation}
  {\cal O}_1 = S_{L/2}^z,\ \quad {\cal O}_2 = S_{L/2}^+ S_{L/2+1}^-\ . 
\end{equation}
Focusing on the Krylov space with $N_d = N_u = 0$, Figs.\ 
\ref{FigS3}~(a)-(f) show the ``cloud'' of diagonal elements $\langle n|{\cal 
O}_{1/2}|n\rangle$ for three different values of the deformation parameter $\nu 
= 0.5,1,2$, and two different system sizes $L = 
10,12$. Generally, the $\langle n|{\cal 
O}_{1/2}|n\rangle$ behave consistent with other known examples in the 
literature \cite{dalessio2016S}, i.e., the distributions are relatively broad 
at 
the edges of the spectrum, while they narrow down in the center. 
Especially for ${\cal O}_1$, we find that the distribution of 
the $\langle n|{\cal 
O}_{1}|n\rangle$ notably depends on $\nu$, with a broader distribution for $\nu 
= 2$ and a narrower distribution for $\nu = 0.5$.   
While it is hard to see from the bare distributions in Fig.\ 
\ref{FigS3}, we have checked that the variances of the $\langle n|{\cal 
O}_{1/2}|n\rangle$ actually decrease approximately 
exponentially with increasing $L$ for all $\nu$ (see insets in 
Fig.\ \ref{FigS3} and caption for description). 
Thus, we expect that in the thermodynamic 
limit $L \to \infty$, the overwhelming majority of eigenstates of ${\cal 
H}_\nu$ 
follows the ETH. 

In Fig.\ \ref{FigS3}, we additionally highlight the expectation value $\langle 
S_\nu|{\cal O}_{1/2}|S_\nu\rangle$ with respect to the exact 
eigenstate $|S_\nu\rangle$. In the case of ${\cal O}_1$, we 
find $\langle 
S_\nu|{\cal O}_{1}|S_\nu\rangle \approx 0$ for $\nu = 0.5,1$ such that 
$\ket{S_\nu}$ is indistinguishable from the thermal eigenstates in its 
vicinity. Interestingly, for $\nu = 2$ [Fig.\ \ref{FigS3}~(e)], $\ket{S_\nu}$ 
yields a nonzero value of ${\cal O}_1$, which also reflects 
itself in the quench dynamics of 
the domain-wall state at $\nu = 2$ (Fig.\ 5 of main text).
In contrast, in the case of ${\cal O}_2$ [Figs.~\ref{FigS3}~(b),(d),(f)], 
we find $\langle 
n|{\cal O}_{2}|n\rangle \approx 0$ for most $\ket{n}$, while 
the exact state $\ket{S_\nu}$ yields a nonzero expectation 
value which is a clear 
outlier well separated from the bulk of the thermal states, demonstrating 
the embedding of a 
quantum many-body scar into the spectrum of ${\cal H}_\nu$. 

\subsection{Additional data on domain-wall melting}
\begin{figure}[tb]
 \centering
 \includegraphics[width=0.9\columnwidth]{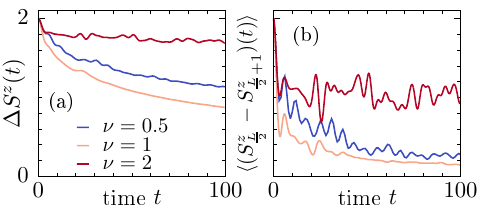}
 \caption{(a) Magnetization difference $\Delta S^z(t)$ [Eq.\ 
\eqref{Eq::DeltaS}] between the two halves of the chain. (b) 
Magnetization difference $\langle S_{L/2}^z(t)\rangle - \langle 
S_{L/2+1}^z(t)\rangle$ between the two central lattice sites. Note that the 
curve for $\nu = 2$ in (b) is very similar to the Loschmidt echo ${\cal 
L}(t)$ in Fig.\ 5. We have $L = 16$ in all cases.}
 \label{FigS4}
\end{figure}

Let us present additional data on the melting of domain-wall initial states 
considered in Fig.\ 5 of the main text. Figure \ref{FigS4}~(a) 
shows the magnetization difference $\Delta S^z(t)$ between the 
two halves of the system for $\nu = 0.5, 1, 2$, 
\begin{equation}\label{Eq::DeltaS}
 \Delta S^z(t) = \frac{2}{L}\sum_{\ell = 1}^{L/2} \left(  \langle 
S_\ell^z(t)\rangle -\langle S_{L/2+\ell}^z(t)\rangle  \right)\ . 
\end{equation}
Consistent with the earlier data shown in Figs.\ 5~(c) and (d), 
$\Delta S^z(t)$ continues to decay for  
$\nu = 0.5, 1$ even at long times $t = 100$, while the dynamics is very slow 
for 
$\nu = 
2$. Moreover, Fig.\ \ref{FigS4}~(b) shows the magnetization 
difference only between the two central sites of the lattice. We find that 
$\langle 
S_{L/2}^z(t)\rangle - \langle 
S_{L/2+1}^z(t)\rangle$ decays almost to zero for $\nu = 0.5,1$ (i.e., the spin 
profile becomes smooth in the center), whereas a finite magnetization 
jump 
remains in the case of $\nu = 2$. 
In this context, it is also instructive to  
compare the curve of $\langle 
S_{L/2}^z(t)\rangle - \langle 
S_{L/2+1}^z(t)\rangle$ for $\nu = 2$ in Fig.\ \ref{FigS4}~(b) to the decay of 
the Loschmidt 
echo ${\cal L}(t)$ in Fig.\ 5~(a) of the main text. Quite remarkably, 
one finds  
that the two curves are very similar to each other, even on the level of 
individual (finite-size) fluctuations. This observation is in good 
agreement with a typicality-based framework developed in 
\cite{Reimann2020S}, which predicts that the dynamics of observables is 
closely related to ${\cal L}(t)$ in situations where one eigenstate (here the 
exact state $\ket{{\cal S}_\nu}$) is 
macroscopically populated.

\subsection{Entanglement growth}
\begin{figure}[tb]
 \centering
 \includegraphics[width=1\columnwidth]{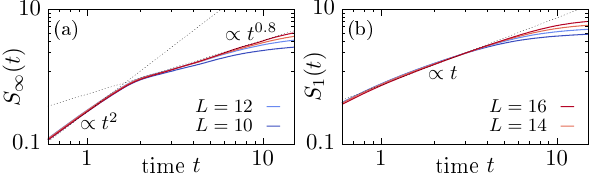}
 \caption{(a) $S_\infty(t)$ and (b) $S_1(t)$ for 
chains with PBC and different $L$, obtained by averaging over multiple 
product states in the $S^z = 0$ sector. Power laws are shown for comparison.}
 \label{FigS5}
\end{figure}

To complement the analysis of anomalous hydrodynamics from the main text, we 
here study the 
R\'enyi 
entropies $S_\alpha(t) = \ln 
\text{Tr}[\rho_A^\alpha]/(1-\alpha)$, $\rho_A = 
\text{Tr}_B \ket{\psi(t)}\bra{\psi(t)}$, which were argued to grow 
subballistically for $\alpha > 1$ \cite{Rakovszky2019S, Huang2020S}. 
Focusing on the extremal case
$S_\infty(t) = -\ln \lambda_\text{max}$, where $\lambda_\text{max}$ is the 
largest eigenvalue of $\rho_A$, ${\cal H}_\nu$ indeed yields a rather 
unusual build-up of 
entanglement with $S_\infty(t) \propto 
t^2$ at short times (similar to strongly coupled holographic systems 
\cite{Liu2014S}) and $S_\infty(t) \propto 
t^{0.8}$ at larger $t$, see Fig.\ \ref{FigS5}~(a). While  
$S_{\infty}(t)$ thus neither grows diffusive \cite{Rakovszky2019S, Huang2020S}, 
nor agrees with the conjecture 
$S_{\alpha>1}(t) \propto t^{1/z}$ \cite{Rakovszky2019S, Znidaric2020S, 
Rakovszky2021S}, we note that models with spin $S \geq 1$ might exhibit 
subtleties \cite{Znidaric2020S, Rakovszky2021S}. In this context, let us stress 
that a numerical analysis is complicated due to  finite-size effects such that 
potential changes in the growth rate at later times cannot be accessed. In 
contrast 
to $S_\infty(t)$, the von Neumann entropy $S_1(t) \propto t$ [Fig.\ 
\ref{FigS5}~(b)] scales linearly as 
expected.

\subsection{Quench dynamics for another initial state}

\begin{figure}[tb]
 \centering
 \includegraphics[width=\columnwidth]{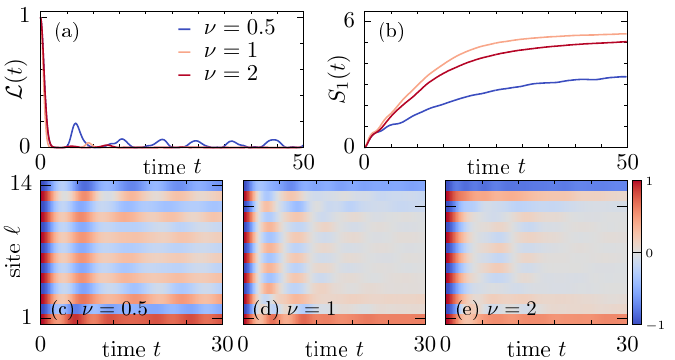}
 \caption{[(a),(b)] ${\cal L}(t) = |\braket{\psi(t)|\psi}|^2$ and 
entanglement $S_1(t) = \text{Tr}[\rho_A(t) \ln \rho_A(t)]$ for the initial 
state $\ket{udud\cdots}$. [(c)-(e)] Color plot of 
$\bra{\psi(t)}S_\ell\ket{\psi(t)}$ for $\nu = 0.5,1,2$. We have $L = 14$,
$c_1 = c_3 = 1$, and $c_2 = -1$ in all cases.}
 \label{FigS6}
\end{figure}

In the main text, we have exemplified the impact of the scar $\ket{{\cal 
S}_\nu}$ on the dynamics by considering a domain-wall initial state $|u\cdots 
ud\cdots d\rangle$, which can be tuned between a localized and a delocalized 
regime depending on the choice of $\nu$. Let us here present additional data 
for another initial state, namely a N\'eel-like state $\ket{\psi} = |udud\cdots 
\rangle$ which likewise belongs to the Krylov space ${\cal K}_{00}$. In 
contrast 
to the domain wall, which maximizes the area ${\cal A}$, the N\'eel state 
yields a much smaller ${\cal A}$. As a consequence, in contrast to Fig.\ 
5 in the main text, the overlap between $\ket{{\cal S}_\nu}$ and 
$\ket{\psi}$ is now enhanced for $\nu < 1$, leading to slower thermalization  
for $\nu< 1$. This is 
demonstrated in Fig.\ \ref{FigS6}, where ${\cal L}(t) = 
|\braket{\psi(t)|\psi}|^2$ quickly decays for $\nu = 1,2$, while revivals are 
present for $\nu = 0.5$ [Fig.\ \ref{FigS6}~(a)], accompanied by a faster growth 
of entanglement for $\nu \geq 1$ [Fig.\ \ref{FigS6}~(b)]. Likewise, the 
$\cdots ud \cdots$ pattern remains more stable for $\nu = 0.5$ [Fig.\ 
\ref{FigS6}~(c)], while fast bulk thermalization occurs for $\nu = 1,2$ [Figs.\ 
\ref{FigS6}~(d) and (e)].

\subsection{Construction of $\ket{{\cal S}_\nu}$}\label{Sec::App_Construct}

According to Eq.\ (3) in the main text, the exact eigenstate 
$\ket{{\cal 
S}_\nu}$ 
is given by the area-weighted superposition of all basis states in a given 
Krylov space. While we 
refer to Refs.\ \cite{Bravyi2012S, Zhang2017S, Movassagh2016S} 
for details on $\ket{{\cal S}_\nu}$ indeed being an eigenstate of 
${\cal H}_\nu$, let us here derive the second part of Eq.\ 
(3), i.e., 
relating the area of a given spin configuration to its dipole 
moment (see also \cite{Langlett2021S}).   
To this end, we introduce the height $h_\ell$ of a spin configuration at 
position $\ell$, 
\begin{equation}\label{Eq::Height}
 h_\ell = \sum_{l = 1}^\ell S_l^z\ ,   
\end{equation}
where we define $h_0 = 0$. 
Using this, the area ${\cal A}_k = \bra{k}{\cal A}\ket{k}$ of a spin 
configuration follows as, 
\begin{align}
 {\cal A} &= \sum_{\ell = 1}^L (h_\ell + h_{\ell-1})/2 \\
    &= \sum_{\ell = 1}^L \left(\sum_{l = 1}^{\ell} S_l^z + \sum_{l=1}^{\ell-1} 
S_l^z\right)/2 \\
&= S^z/2 + \sum_{\ell = 1}^L \sum_{l=1}^{\ell-1} S_l^z  \\
&=  S^z/2 + \sum_{\ell = 1}^L (L - 
\ell)S_\ell^z \\
&= (2L+1)S^z/2 -{\cal P}\ ,   \label{Eq::LastE}
\end{align}
where $S^z = \sum_{\ell=1}^L 
S_\ell^z$ and ${\cal P}$ is the dipole operator. Since the $\ket{{\cal S}_\nu}$ 
are defined 
within a sector with fixed magnetization $S^z$, 
$\bra{k}S^z\ket{k}$ will be independent of the specific basis state $\ket{k}$ 
such that the first term on the right hand side of Eq.~\eqref{Eq::LastE} can 
be dropped. It then follows that,  
\begin{equation}
 \ket{{\cal S}_\nu} = \frac{1}{\sqrt{M_\nu^\prime}} \sum_{k} \nu^{{\cal 
A}_k} \ket{k} 
= \frac{1}{\sqrt{M_\nu}} \sum_{k} \nu^{-{\cal P}_k} \ket{k}\ . 
\end{equation}
where $M_\nu^\prime = \sum_k \nu^{2{\cal A}_k}$ and $M_\nu = \sum_k 
\nu^{-2{\cal P}_k}$. 

\subsection{Dynamical quantum typicality}\label{App:Typ}

The correlation function $C(r,t)$ in Eq.\ (4) can be 
efficiently calculated by means of the concept of dynamical quantum typicality 
(DQT) \cite{Jin2021S, Heitmann2020S}. To this end, let $\ket{\cal R} = \sum_k 
c_k 
\ket{k}$ be a Haar-random state drawn from the full Hilbert space, i.e., in 
practice the sum runs over the $3^L$ computational basis states $\ket{k}$ and 
the real and imaginary parts of the complex coefficients $c_k$ are drawn from a 
Gaussian distribution with zero mean. We assume $\sum_k |c_k|^2 = 1$. According 
to 
DQT, $C(r,t)$ can then be approximated as (see also \cite{Richter2021S, 
Chiaracane2021S} for detailed derivations), 
\begin{equation}\label{Eq::DQT}
 C(r,t) = \bra{{\cal R}_\ell(t)}S_{\ell+r}^z\ket{{\cal R}_\ell(t)} + 
\varepsilon(\ket{{\cal R}})\ , 
\end{equation}
where $\ket{{\cal R}_\ell} = \sqrt{S_\ell^z + 1}\ket{{\cal R}}$ and $\ket{{\cal 
R}_\ell(t)} = e^{-i{\cal H}t} \ket{{\cal R}_\ell}$. Importantly, the 
statistical error of the approximation in 
Eq.\ \eqref{Eq::DQT} scales as $\varepsilon(\ket{{\cal R}}) \propto 
1/\sqrt{3^{L}}$ \cite{Jin2021S, Heitmann2020S}, and can therefore be neglected 
already for intermediate system sizes. Thus, the correlation function $C(r,t)$ 
is faithfully approximated by the expectation value of $S_{\ell+r}^z$ within 
the 
random state $\ket{{\cal 
R}_\ell(t)}$. Since the time evolution of $\ket{{\cal 
R}_\ell(t)}$ can be evaluated efficiently by standard sparse-matrix techniques 
\cite{Fehske2009S}, 
$C(r,t)$ can be 
simulated for system sizes beyond the range of ED.   

\subsection{Details on stochastic cellular automaton dynamics}\label{App::CA}

In order to substantiate our direct simulations of the transport properties of 
the Motzkin chain ${\cal H}_\nu$ (i.e., as obtained under full quantum 
evolution 
on system sizes $L \leq 18$), we have constructed a stochastic cellular 
automaton (CA) circuit (or Markov chain) \cite{Iaconis2019S, 
Feldmeier2020S, Morningstar2020S, Medenjak2017S, Gopalakrishnan2018S}.  
The CA circuit is constructed in such a way, that it mimics the terms appearing 
in ${\cal H}_\nu$ [Fig.\ 1~(a)]. It is composed of two-site 
updates ${\cal U}$ which map 
product states from the $3^L$-dimensional 
computational basis to other product states (e.g., ${\cal U}\ket{0du00d\cdots} 
\to \ket{d00u0d\cdots}$), i.e., no entanglement is created when the initial 
state of the circuit is itself a member of the computational basis. As a 
consequence, classical simulations of large system sizes and long time scales 
are possible \cite{Iaconis2019S, 
Feldmeier2020S, Morningstar2020S, Medenjak2017S, Gopalakrishnan2018S}. 
and 

In Fig.\ 3 of the main text, a single exemplary time step of the 
stochastic automaton 
evolution is illustrated. Starting with an arbitrary configuration in the 
computational basis, we consider three different types of local updates, named 
$D$, $U$, and $V$ [in accordance with the projectors in Eq.\ 
(1)]. Given a configuration of the two 
neighboring spins, the appropriate local update is chosen, where $D$ acts 
as $|0d\rangle \leftrightarrow |d0\rangle$, $U$ acts 
as $|0u\rangle \leftrightarrow |u0\rangle$, and $V$ acts as $|00\rangle 
\leftrightarrow |ud\rangle$, i.e., these updates correspond to the off-diagonal 
terms of ${\cal H}_\nu$. However, with a probability of $1/2$ (hence {\it 
stochastic} circuit), the updates $D$, $U$, or $V$ are replaced by $\bar{D}$, 
$\bar{U}$, $\bar{V}$, which act as the identity, such that the local 
configuration remains unchanged (see gray gates in Fig.\ 3), i.e., 
these cases correspond to the diagonal terms of ${\cal H}_\nu$.    
Likewise, if the two-site configuration is given by $|dd\rangle$, 
$|uu\rangle$, 
or $|du\rangle$, no local update is performed as ${\cal H}_\nu$ does not 
contain 
corresponding terms (such a case is not shown in 
Fig.\ 3).
A full time step in the circuit then consists of two layers of local two-site 
updates, where the local unitary transformations first act on all even bonds 
and subsequently on all odd bonds.
Moreover, the infinite-temperature spin-spin correlation function $C(r,t)$ in 
Eq.\ 
(4) is obtained by averaging the 
classical quantity $S_{\ell+r}^z(t)S_{\ell}^z(0)$ over sufficiently many 
initial 
spin configurations. 

\subsection{Derivation of spin-current operator}\label{App::SCur}

Let us derive the expression for the spin-current operator of the Motzkin 
chain. Inspecting the Hamiltonian in Eq.\ (1), the 
relevant 
contributions to spin transport are given by the off-diagonal terms 
$c_1\nu(-\ket{0d}\bra{d0}-\ket{d0}\bra{0d})/(1+\nu^2)$, 
$c_2\nu(-\ket{0u}\bra{u0}-\ket{u0}\bra{0u})/(1+\nu^2)$, and 
$c_3\nu(-\ket{00}\bra{ud} - 
\ket{ud}\bra{00})/(1+\nu^2)$, while the diagonal terms can be ignored. It is 
further
helpful to rewrite the above expressions in terms of spin-$1$ operators. For 
example, one finds \cite{Chen2017S}, 
\begin{equation}
 \ket{0d}\bra{d0} = S_1^+ S_1^z S_2^z S_2^-\ ,  
\end{equation}
and the other off-diagonal terms have similar representations that we here 
omit for brevity. The spin current now follows from the lattice continuity 
equation $\tfrac{d}{dt} S_\ell^z = i[{\cal H}_\nu,S_\ell^z] = 
j_{\ell-1}-j_{\ell}$ \cite{Bertini2021S}. 
Note that the commutator is non-vanishing only for the local terms of ${\cal 
H}_\nu$ acting on 
sites 
$\ell-1,\ell$ and $\ell,\ell+1$. Using the identity 
$[S_\ell^\pm,S_{\ell'}^z] = \mp S_\ell^\pm \delta_{\ell\ell'}$ and carrying out 
some straightforward manipulations, one finds that $j = \sum_\ell j_\ell$ with, 
\begin{align}\label{Eq::FullFullCurr}
 j_\ell = \frac{i\nu}{1+\nu^2}&\big[(c_3 S_\ell^z S_\ell^+ - c_1 S_\ell^+ 
S_\ell^z)S_{\ell+1}^z S_{\ell+1}^- \\
        &+S_{\ell}^- S_\ell^z (c_2 S_{\ell+1}^z S_{\ell+1}^+ - c_3 
S_{\ell+1}^+ S_{\ell+1}^z) \nonumber \\
&+ S_\ell^z(c_1S_{\ell}^- S_{\ell+1}^+ - c_2 S_{\ell}^+ S_{\ell+1}^-) 
S_{\ell+1}^z\big]\ . \nonumber 
\end{align}

At $\nu = c_i = 1$, the local terms in Eq.\ \eqref{Eq::FullFullCurr} can be 
substantially simplified and take on the form, 
\begin{align}\label{Eq::FullCurr}
 j_\ell = \frac{i}{2}&\big[S_\ell^+ S_{\ell+1}^z S_{\ell+1}^- + 
S_\ell^- 
S_\ell^z S_{\ell+1}^+  \\
& + S_\ell^z (S_\ell^- S_{\ell+1}^+ - S_\ell^+ S_{\ell+1}^-)S_{\ell+1}^z\big]\ 
,  \nonumber
\end{align}
where we again exploited the commutator relations of $S^\pm_\ell$ and 
$S_\ell^z$. It is insightful to contrast Eq.\ \eqref{Eq::FullCurr} with 
the 
spin current of a more common spin-$1$ model, e.g, the Heisenberg chain, ${\cal 
H}_\text{Heis} = \sum_\ell {\bf S}_\ell \cdot {\bf S}_{\ell+1}$, for which the 
current has the well-known form 
$j_\text{Heis} = \tfrac{i}{2}\sum_\ell (S_\ell^+ S_{\ell+1}^- - S_\ell^- 
S_{\ell+1}^+)$ \cite{Bertini2021S}. Comparing $j_\text{Heis}$ to Eq.\ 
\eqref{Eq::FullCurr}, $j$ is essentially a dressed version of 
$j_\text{Heis}$, which formalizes that some configurations do not contribute 
to the transport of spin in the Motzkin chain.


\begin{thebibliography}{99}
\bibitem{NoteCi}
With the exception of Fig.\ \ref{Fig4}, we choose $c_1 = 
c_3 = 1$ and 
$c_2 = -1$ throughout this work, for which the $\ket{{\cal S}_\nu}$ are located 
close to 
the center of the spectrum.
\bibitem{NotePos}
Note that the 
specific site $\ell$ is irrelevant in the case of PBC, while for OBC, we 
use $\ell = L/2$.
\bibitem{NoteEntang}
We refer to the R\'enyi entropies 
$S_\alpha(t) = \ln 
\text{Tr}[\rho_A^\alpha]/(1-\alpha)$ which yield $S_1(t) = 
-\text{Tr}[\rho_A \ln \rho_A]$ for $\alpha = 1$.
\end{thebibliography}
 \end{document}